# Tunable quantum anomalous Hall octet driven by orbital magnetism in bilayer graphene

**Authors:** Fabian R. Geisenhof[1], Felix Winterer[1], Anna M. Seiler[1], Jakob Lenz[1], Tianyi Xu[2], Fan Zhang[2]*, R. Thomas Weitz[1,3,4,5]*

**Affiliations:**

[1]Physics of Nanosystems, Department of Physics, Ludwig-Maximilians-Universität München, Geschwister-Scholl-Platz 1, Munich 80539, Germany

[2]Department of Physics, University of Texas at Dallas, Richardson, TX, 75080, USA

[3]Center for Nanoscience (CeNS), Schellingstrasse 4, Munich 80799, Germany

[4]Munich Center for Quantum Science and Technology (MCQST), Schellingstrasse 4, Munich 80799, Germany

[5]1st Physical Institute, Faculty of Physics, University of Göttingen, Friedrich-Hund-Platz 1, Göttingen 37077, Germany

*Corresponding authors.

Email: zhang@utdallas.edu, thomas.weitz@uni-goettingen.de

**Abstract:**

The quantum anomalous Hall (QAH) effect – a macroscopic manifestation of chiral band topology at zero magnetic field – has only been experimentally realized by magnetic doping of topological insulators[1–3] and delicate design of moiré heterostructures[4–8]. However, the seemingly simple bilayer graphene without magnetic doping or moiré engineering has long been predicted to host competing ordered states with QAH effects[9–11]. Here, we explore states in bilayer graphene with conductance of 2 $e^2/h$ that not only survive down to anomalously small magnetic fields and up to temperatures of 5 K, but also exhibit magnetic hysteresis. Together, the experimental signatures provide compelling evidence for orbital magnetism driven QAH behavior with a Chern number tunable via electric and magnetic fields as well as carrier sign. The observed octet of QAH phases is distinct from previous observations due to its peculiar ferrimagnetic and ferrielectric order that is characterized by quantized anomalous charge, spin, valley, and spin-valley Hall behavior.

**Main Text:**

Intricate interplay between single-particle effects such as the band topology and many-body effects such as the electron-electron interaction determines the electronic ground states of many low-dimensional systems. An especially interesting novel class is given by systems in which quasiparticle Berry curvature gives rise to orbital instead of spin magnetic moments with the consequence that effects usually requiring substantial spin-orbit coupling and/or intentional magnetic doping can occur spontaneously. A prominent example is the quantum anomalous Hall (QAH) effect that displays quantized Hall resistance at zero magnetic field due to the presence of orbital magnetic order. This QAH phase is characterized by a finite number of topologically protected chiral edge channels. So far, it has been experimentally realized in two distinct types of systems. In magnetically doped topological insulators[1–3], spin-orbit coupling inducing topological properties and aligned magnetic dopants breaking time-reversal symmetry lead to topologically non-trivial Chern bands[12,13]. In these spin Chern insulators magnetism occurs mainly due to ordering of electron spin moments. However, a Chern insulator can also emerge solely due to a spontaneous polarization of the orbital magnetic moments[9,10,14], as it has recently been observed in delicately designed moiré heterostructures[4–8]. In these orbital Chern insulators, orbital magnetism arises because of spontaneous gap opening in the half-filled quasiparticle Dirac bands[9–11,14]. Gapped Dirac bands with nontrivial Berry curvature induced orbital magnetic moments have also been predicted[9,15] and observed in naturally occurring purely carbon based systems such as bilayer graphene[16,17] and its rhombohedral cousins[18]. However, orbital magnetism has not been clearly pinpointed experimentally in such a simple system as pure bilayer graphene, despite theoretical studies[9–11] predicting that some of the competing ground states should exhibit non-vanishing exchange interaction-driven quantized Hall conductivities at zero magnetic field.

Here, we report the observation of $\nu = \pm 2$ states at anomalously small magnetic fields of $\sim$ 20 mT in suspended dually-gated bilayer graphene devices. Additionally, we observe magnetic hysteresis, which strongly evidences that the $\nu = \pm 2$ states are QAH phases driven by orbital magnetism in pure bilayer graphene.

To fabricate suspended dually-gated bilayer graphene devices suitable flakes are identified with optical microscopy and examined using apertureless scanning near-field imaging (s-SNOM), which allows to select flakes free of electronically-active domain walls[19–21] that might mask fragile quantum Hall phases. Multiple steps of standard electron-beam lithography and subsequent wet etching were performed subsequently to prepare devices[22]. A false color scanning electron microscope image of a typical device is shown in Fig. 1a. To improve sample cleanliness current annealing was employed directly in the cryostat[22,23]. Unless stated otherwise, the transport data was taken using standard lock-in techniques and an ac bias current of 1 - 10 nA at a temperature of $T$ < 10 mK.

Sweeping both top and bottom gate voltages, $V_t$ and $V_b$, at zero magnetic field yields the well-known map of the differential conductance including the interlayer electric-field induced insulating states as well as the exchange interaction-induced gapped phase near zero electric field (Fig. 1b)[24–26]. The observation of the latter and the location of the charge neutrality point at $V_t \approx V_b \approx 0$ demonstrates the high quality of the device. The dual-gate structure allows independent tunability of the charge carrier density $n$ and the perpendicular electric field $E_\perp$. Sweeping $n$ at $E_\perp = 0$ reveals a residual charge density inhomogeneity of < $10^9$ cm$^{-2}$ (Fig. 1b, inset), underlining the high quality of the device[27].

Varying both $n$ and $E_\perp$ while applying a perpendicular magnetic field of $B = 3$ T (Fig. 1c) reveals the lowest quantum Hall plateaus in bilayer graphene with the integer filling factors ranging from $\nu$ = -4 to 4[24,28–32], resulting from the spontaneous symmetry breaking in the anomalous $N = 0$ Landau level octet. The emergence of the quantum Hall plateaus also allows us to calculate and subtract the contact resistance[22]. As identified previously, only the $\nu = 0$ and ±4 quantum Hall states are resolved at $E_\perp = 0$ (Fig. 1c), whereas the $\nu = \pm1$, ±2 and ±3 states emerge only at a larger finite electric field[24,32,33]. At a lower magnetic field of $B = 0.8$ T (Fig. 1d) only the $\nu = 0$, ±2 and ±4 states emerge. Surprisingly, in contrast to the $\nu = \pm4$ states, the $\nu = \pm2$ states are only stable in an intermediate range of positive or negative electric field (in the four green-colored regions in Fig. 1d), i.e. both larger and smaller electric fields can destabilize the $\nu = \pm2$ states.

While the $\nu = \pm2$ states in bilayer graphene have been observed previously at B > 1.2 T[24,32], their exact nature especially with lowering magnetic field toward the $B = 0$ limit where one can expect intricate QAH phases and phase transitions as function of electric field has not been identified previously. The order parameters of these states are particularly interesting, since they can unveil the yet unclear ground state of bilayer graphene in the $B = 0$ limit[9–11,34]. Due to the quadratic band touching and nontrivial winding numbers, the exchange interaction in bilayer graphene is peculiarly strong and produces nontrivial quasiparticle topological properties[10]; various symmetry-broken states have been suggested as gapped competing ground states[9,11,22], with two families of QAH phases exhibiting orbital magnetism[9]. One family manifesting Hall conductance of 4 $e^2/h$, simply termed QAH phase, is a bilayer counterpart of the Haldane QAH state, in which electrons from different valleys spontaneously polarize to different layers resulting in a $Z_2$ orbital ferromagnetic order. The other family manifesting Hall conductance of 2 $e^2/h$, termed "ALL" phase, is even more exotic; as quasiparticles of one spin can choose either one of the two quantum valley Hall (QVH) phases – electrons from both valleys polarize to the same layer resulting in a $Z_2$ ferroelectric order – whereas quasiparticles of the other spin can choose either one of the two aforementioned QAH phases (see Fig. 1e and Fig. S2). In total, there are eight different ALL phases[22] with Chern number $C = \pm 2$ or Hall conductance $\sigma^{(CH)} = \pm 2$ $e^2/h$. Markedly, this ALL phase exhibits quantized anomalous charge, spin, valley, and spin-valley Hall effects[9,10]. Because of its partial layer polarization, the ALL phase can be stabilized with an interlayer electric field, which fits well to our observations. At very high electric fields, the phase vanishes again, losing stability against a fully layer polarized QVH phase. Furthermore, applying a perpendicular magnetic field should lower its energy since the field can couple to the quasiparticle orbital magnetic moments[9,10]. Therefore, switching the sign(s) of the applied $n$, $E_\perp$, and/or $B$ results in a quantum phase transition between two different ALL phases, as sketched in Fig. 1f. By flipping $E_\perp$ and $n$, the layer polarization of the QVH species and the orbital magnetization of the QAH species become opposite, respectively. By flipping $B$ both the orbital magnetization and the spin of the QAH species become opposite. Comparing the measurement at $B = 3$ T (Fig. 1c) and $B = 0.8$ T (Fig. 1d), we find that the electric field range at which the ALL phase emerges at $B = 3$ T expands towards higher electric fields. This demonstrates the enhanced stability of the $\nu = \pm 2$ QAH phases with increasing magnetic field. More data on the evolution of these phases in the magnetic field is shown in Fig. S4.

So far we have examined the stability of the $\nu = \pm 2$ QAH phases at small but finite magnetic fields. Since these phases are driven by the exchange interaction-induced orbital magnetism, they should, however, also be stable towards $B = 0$. To this end, we have recorded multiple fan diagrams around $B = 0$ at various electric fields (Fig. 2). From the fan diagram at $E_\perp = -20$ mV/nm (Fig. 2a,b) we can see that both the $\nu = \pm 2$ and $\pm 4$ states emerge already at unusually small magnetic field. We focus here on the $\nu = \pm 2$ QAH phases, since they previously escaped observation at such low magnetic fields[30,32], and because they are the most exotic ones among the competing ground states of bilayer graphene at $B = 0$: quasiparticles of one spin form a QVH phase whereas those of the other spin a QAH phase[9–11], see Fig. 1e, S2 and S3. Carefully examining the derivative of the conductance (Fig. 2b) to track fluctuations near incompressible quantum states provides more insight, since the traceable fluctuations are assignable to specific filling factors using their slopes and can appear even before the corresponding quantum Hall states emerge in conventional magneto-transport measurements[35]. Investigating the derivative of the differential conductance at various electric fields (Fig. 2b,c) demonstrates that both the $\nu = \pm 2$ and $\pm 4$ states already emerge at magnetic fields well below $B = 100$ mT, but that they differ in their electric field dependences. While the amount of fluctuations at finite $B$ corresponding to the $\nu = \pm 4$ states decreases with increasing negative electric field (Fig. 2c, left panels), the $\nu = \pm 2$ states are prominent at $E_\perp = -15$ mV/nm but disappear at zero and high negative electric fields. Additionally, high-resolution scans around zero magnetic field (Fig. 2c, right panels) reveal that the $\nu = \pm 2$ states are also present for $B < 100$ mT. In fact, they do persist to $B < 20$ mT, which is even further than the $\nu = \pm 4$ states. This provides strong evidence that the QAH $\nu = \pm 2$ phases are potential ground states of bilayer graphene at $B = 0$ in addition to the previously identified $\nu = 0$ layer antiferromagnet (LAF) and $\nu = \pm 4$ QAH phases[24–26]. The observation that the $\nu = \pm 2$ states can be stabilized by a combination of $B$ and $E_\perp$ fields is consistent with their partial layer polarizations and orbital magnetic characters. Finally, for very high electric fields, fluctuations with an infinite slope that trace the fully layer polarized $\nu = 0$ QVH phase dominate the fan diagram. More data showing fan diagrams at additional electric fields are displayed in Fig. S5.

While the electric and magnetic field dependences and the stability down to $B \sim 0$ T support the presence of the QAH phases, we have looked for more direct proof of the presence of their orbital magnetism. Most prominently, magnetism can be shown by hysteretic behavior of resistance – whereas in our two-terminal measurements, the absolute contributions of both longitudinal resistance $R_{xx}$ and Hall resistance $R_{xy}$ are measured simultaneously. Indeed, by sweeping the magnetic field at constant $\nu = -2$ and $E_\perp = -17$ mV/nm, we have recorded a magnetic hysteresis (Fig. 3a). Forward and reverse sweeps are mirror symmetric with respect to the $B = 0$ line, with the hysteretic behavior starting at about $B = \pm 650$ mT. Additionally, the hysteresis is highly reproducible upon repeated sweeps and we also observe it in a second device (Fig. S7). This hysteretic behavior provides compelling evidence for the emergence of magnetism in pure bilayer graphene, which has not been observed previously in a moiréless purely carbon-based 2D system. Given the vanishing spin-orbit coupling in bilayer graphene, the magnetism is primarily of orbital nature, which stems from the opposite mean-field gaps in the two valleys in one of the two spin species[9–11], see Fig. 1e, S2 and S3.

The intimate relation of the orbital magnetism to the $\nu = \pm 2$ QAH phases can be further validated by a series of test measurements. First, cyclic $B$ sweeps for fixed $n$ (and consequently varying ν) do not show hysteretic behavior (Fig 3b). These measurements were performed at $n$ = -0.25, -0.5, -1.0 × $10^{11}$ cm$^{-2}$ corresponding to the quantum Hall states of ν = -1, -2, -4 at $B$ = 1 T, respectively. This implies that when the magnetic field is swept toward $B$ = 0, the sample reaches quantum Hall states with higher filling factors up to ν = -12 for $n$ = -1.0 × $10^{11}$ cm$^{-2}$, far away from the ν = -2 QAH phase.

A second set of test measurements addresses the electric field dependence in the region where the ν = -2 QAH state is stable (Fig. 3c). Consistently, at $E_\perp$ = 0 we do not observe any hysteretic behavior since a ν = -2 state is not observable here. At $E_\perp$ = -10 mV/nm, in agreement with the observations from the fan diagrams (Fig. 2), hysteretic behavior starts to emerge, and the hysteresis loop area reaches its maximum at $E_\perp$ = -17 mV/nm. With increasing negative electric field, the hysteresis decreases again and vanishes completely at $E_\perp$ = -60 mV/nm, where the fully layer polarized ν = 0 QVH phase dominates. These observations are consistent with the electric field dependence of the ν = -2 state in Fig. 2 and the partial layer polarization of the ν = -2 QAH phase in Fig. 1e.

Finally, the hysteretic behavior vanishes at constant finite electric field if the filling factor is detuned significantly away from ν = -2 (Fig. 3d). Since the quantum Hall states with ν = -1 and -3 do not emerge at $B$ < 1 T, all nominal fillings in the range of -1 < ν < -3 correspond to the ν = -2 state. In this range, we observe hysteresis with the loop area reaching its maximum at ν = -2.5. When setting the filling factor to higher or lower nominal filling, e.g. ν = -1 or ν = -3, the hysteresis vanishes.

As a final test of the stability of the ν = -2 QAH phase, we have investigated its temperature dependence at various electric fields at $B$ = 0.5 T (Fig. 4). The full temperature dependent transport data is shown in Fig. S6. While a quantitative estimation of the bulk gap in the $\nu = -2$ state via calculation of its activation energy $\Delta_{\nu=-2}$ is challenging due to the potential presence of disorder, we use these estimates of a relative judgement of the stability of the various observed phases. Fig. 4d shows an Arrhenius plot of the conductance at $n(\nu=-2)$ and various electric fields. Since the temperature dependence of the conductance[24] follows $\sigma \sim exp(-\Delta_\nu/(2k_bT))$, in the semi-log graph we can use a linear fit to calculate the energy gap. At zero electric field (Fig. 4a), the $\nu = -2$ state does not persist to $B = 0.5$ T as we have seen in the fan diagrams, and consequently the temperature dependence is very small, indicating a vanishing energy gap. By contrast, at a finite electric field of $E_\perp = 15$ mV/nm there is an evident temperature dependence (Fig. 4b) with an energy gap of $\Delta_{\nu = -2} = (0.09 \pm 0.02)$ meV. Applying an even higher electric field of $E_\perp = 50$ mV/nm (Fig. 4c), the $\nu = -2$ state becomes less stable with a smaller $\Delta_{\nu = -2} = (0.039 \pm 0.001)$ meV, again consistent with its predicted partial layer polarization. We would like to point out that gap energies measured by activation can only give a lower bound for the real gap due to the presence of local disorder[22] but their absolute magnitude can be put into perspective by comparing them with the gaps of the $\nu = \pm 4$ and $\nu = 0$ states as functions of electric field, as shown in Fig. 4e. The behavior of the $\nu = 0$ state with a large gap of $\Delta_{\nu = 0} = 3$ meV at zero electric field, a vanishing gap for an intermediate electric field, and a reappearance for high electric field is consistent with the observation of the phase transition from the interaction-driven layer-balanced gapped LAF phase to the electric field-induced fully layer-polarized gapped QVH state[24,25]. The activation gaps of the $\nu = -4$ and $\nu = -2$ states exhibit very different electric field dependencies but rather similar magnitudes, with $\Delta_{\nu = -4} = (0.08 \pm 0.04)$ meV at $E_\perp =$ 0 mV/nm and $\Delta_{\nu = -2}$ = (0.09 ± 0.02) meV at $E_\perp$ = 15 mV/nm. This observation is surprising, since in previous experiments $\Delta_{\nu = -4} > \Delta_{\nu = -2}$ has been found[30,32,36]. While these previous measurements of the $\nu = \pm 2$ and $\nu = \pm 4$ states were performed at larger magnetic fields or without an independent control of $E_\perp$ and *n,* where the QAH $\nu = \pm 2$ phases may be unstable, the surprising robustness of the $\nu = \pm 2$ states evidenced by the larger activation gaps arises from the electric field coupling to the layer polarization and the magnetic field coupling to the orbital magnetic moments of the quasiparticles.

To conclude, we have tracked the quantum Hall states down to ultra-low magnetic fields and observed compelling evidence that the $\nu = \pm 2$ QAH phases are two competing ground states of bilayer graphene, stabilized when applying a finite interlayer electric field to couple their ferrielectric partial layer polarization. Furthermore, these states exhibit magnetic hysteresis revealing their ferrimagnetic orbital magnetism in pure bilayer graphene. Future measurements using a four-terminal geometry could distinguish between longitudinal and Hall resistances and determine possible switching mechanism of the exotic ordering of such $\nu = \pm 2$ QAH phases by using both magnetic and electric fields. Since the phases persist up to $T \approx$ 5 K, applications in low-dissipation electronics or quantum information science[37] could be exciting further developments.

**Methods:**

Device fabrication: The graphene flakes were exfoliated from a highly ordered pyrolytic graphite (HOPG) block onto Si/SiO$_2$ substrates. Using optical microscopy suitable bilayer flakes are preselected by examining the optical contrast. The flakes were scanned with near-field optical microscopy in order to avoid any structural domain walls within the channel[19,21]. The electrodes (Cr/Au, 5/100 nm), top gate (Cr/Au, 5nm/160 nm) and spacer ($SiO_2$, 140 nm) were fabricated by multiple steps of standard lithography techniques and electron beam evaporation. In order to suspend both the top gates and the bilayer graphene flakes hydrofluoric acid was subsequently used to etch about 150 – 200 nm of the $SiO_2$. After loading the suspended dual-gated bilayer graphene devices in a dilution refrigerator, several cycles of current annealing at 1.6 K were performed (Fig. S1).

Electrical transport measurements: The two-terminal conductance measurements were carried out in a dilution refrigerator with a base temperature of 7 mK. The measurements were performed with an AC bias current of 0.1 – 10 nA at 78 Hz using Stanford Research Systems SR865A and SR830 lock-in amplifiers. Gate voltages were applied using multiple Keithley 2450 SourceMeters. Several home built low-pass RC filters were used in series to reduce high frequency noise.

# References

1. Chang, C.-Z. *et al.* Experimental observation of the quantum anomalous Hall effect in a magnetic topological insulator. *Science* **340,** 167–170 (2013).

2. Tenasini, G. *et al.* Giant anomalous Hall effect in quasi-two-dimensional layered antiferromagnet Co1/3NbS2. *Phys. Rev. Res.* **2** (2020).

3. Zhao, Y.-F. *et al.* Tuning the Chern number in quantum anomalous Hall insulators. *Nature* **588,** 419–423 (2020).

4. Tschirhart, C. L. *et al.* Imaging orbital ferromagnetism in a moiré Chern insulator, Preprint at http://arxiv.org/pdf/2006.08053v1 (2020).

5. Sharpe, A. L. *et al.* Emergent ferromagnetism near three-quarters filling in twisted bilayer graphene. *Science* **365,** 605–608 (2019).

6. Serlin, M. *et al.* Intrinsic quantized anomalous Hall effect in a moiré heterostructure. *Science* **367,** 900–903 (2020).

7. Polshyn, H. *et al.* Electrical switching of magnetic order in an orbital Chern insulator. *Nature* **588,** 66–70 (2020).

8. Chen, G. *et al.* Tunable correlated Chern insulator and ferromagnetism in a moiré superlattice. *Nature* **579,** 56–61 (2020).

9. Zhang, F., Jung, J., Fiete, G. A., Niu, Q. & MacDonald, A. H. Spontaneous quantum Hall states in chirally stacked few-layer graphene systems. *Phys. Rev. Lett.* **106,** 156801 (2011).

10. Zhang, F. Spontaneous chiral symmetry breaking in bilayer graphene. *Synth. Met.* **210,** 9–18 (2015).

11. Nandkishore, R. & Levitov, L. Quantum anomalous Hall state in bilayer graphene. *Phys. Rev. B* **82** (2010).

12. Yu, R. *et al.* Quantized anomalous Hall effect in magnetic topological insulators. *Science* **329,** 61–64 (2010).

13. Zhang, F., Kane, C. L. & Mele, E. J. Surface state magnetization and chiral edge states on topological insulators. *Phys. Rev. Lett.* **110,** 46404 (2013).

14. Zhu, J., Su, J.-J. & MacDonald, A. H. Voltage-Controlled Magnetic Reversal in Orbital Chern Insulators. *Phys. Rev. Lett.* **125,** 227702 (2020).

15. Di Xiao, Yao, W. & Niu, Q. Valley-contrasting physics in graphene: magnetic moment and topological transport. *Phys. Rev. Lett.* **99,** 236809 (2007).

16. Lee, Y. *et al.* Tunable Valley Splitting due to Topological Orbital Magnetic Moment in Bilayer Graphene Quantum Point Contacts. *Phys. Rev. Lett.* **124,** 126802 (2020).

17. Ju, L. *et al.* Tunable excitons in bilayer graphene. *Science* **358,** 907–910 (2017).

18. Shi, Y. *et al.* Electronic phase separation in multilayer rhombohedral graphite. *Nature* **584,** 210–214 (2020).

19. Jiang, L. *et al.* Soliton-Dependent Plasmon Reflection at Bilayer Graphene Domain Walls. *Nat. Mater.* **15,** 840–844 (2016).

20. Ju, L. *et al.* Topological Valley Transport at Bilayer Graphene Domain Walls. *Nature* **520,** 650–655 (2015).

21. Geisenhof, F. R. *et al.* Anisotropic Strain-Induced Soliton Movement Changes Stacking Order and Band Structure of Graphene Multilayers. Implications for Charge Transport. *ACS Appl. Nano Mater.* **2,** 6067–6075 (2019).

22. See supplementary materials.

23. Moser, J., Barreiro, A. & Bachtold, A. Current-induced cleaning of graphene. *Appl. Phys. Lett.* **91,** 163513 (2007).

24. Weitz, R. T., Allen, M. T., Feldman, B. E., Martin, J. & Yacoby, A. Broken-Symmetry States in Doubly Gated Suspended Bilayer Graphene. *Science* **330,** 812–816 (2010).

25. Velasco, J. *et al.* Transport spectroscopy of symmetry-broken insulating states in bilayer graphene. *Nat. Nanotechnol.* **7,** 156–160 (2012).

26. Freitag, F., Trbovic, J., Weiss, M. & Schönenberger, C. Spontaneously gapped ground state in suspended bilayer graphene. *Phys. Rev. Lett.* **108,** 76602 (2012).

27. Nam, Y., Ki, D.-K., Soler-Delgado, D. & Morpurgo, A. F. Electron–hole collision limited transport in charge-neutral bilayer graphene. *Nat. Phys.* **13,** 1207–1214 (2017).

28. Zhao, Y., Cadden-Zimansky, P., Jiang, Z. & Kim, P. Symmetry breaking in the zero-energy Landau level in bilayer graphene. *Phys. Rev. Lett.* **104,** 66801 (2010).

29. Li, J., Tupikov, Y., Watanabe, K., Taniguchi, T. & Zhu, J. Effective Landau Level Diagram of Bilayer Graphene. *Phys. Rev. Lett.* **120,** 47701 (2018).

30. Martin, J., Feldman, B. E., Weitz, R. T., Allen, M. T. & Yacoby, A. Local Compressibility Measurements of Correlated States in Suspended Bilayer Graphene. *Phys. Rev. Lett.* **105,** 256806 (2010).

31. Lee, K. *et al.* Bilayer graphene. Chemical potential and quantum Hall ferromagnetism in bilayer graphene. *Science* **345,** 58–61 (2014).

32. Velasco Jr, J. *et al.* Competing ordered states with filling factor two in bilayer graphene. *Nat. Commun.* **5,** 1–5 (2014).

33. Shi, Y. *et al.* Energy Gaps and Layer Polarization of Integer and Fractional Quantum Hall States in Bilayer Graphene. *Phys. Rev. Lett.* **116,** 56601 (2016).

34. Zhang, J., Nandkishore, R. & Rossi, E. Disorder-tuned selection of order in bilayer graphene. *Phys. Rev. B* **91,** 1–6 (2015).

35. Lee, D. S., Skákalová, V., Weitz, R. T., Klitzing, K. von & Smet, J. H. Transconductance fluctuations as a probe for interaction-induced quantum Hall states in graphene. *Phys. Rev. Lett.* **109,** 56602 (2012).

36. Velasco, J. *et al.* Transport measurement of Landau level gaps in bilayer graphene with layer polarization control. *Nano Lett.* **14,** 1324–1328 (2014).

37. Lian, B., Sun, X.-Q., Vaezi, A., Qi, X.-L. & Zhang, S.-C. Topological quantum computation based on chiral Majorana fermions. *Proc. Natl. Acad. Sci. USA* **115,** 10938–10942 (2018).

**Funding:** R.T.W. and F.R.G. acknowledge funding from the Center for Nanoscience (CeNS) and by the Deutsche Forschungsgemeinschaft (DFG, German Research Foundation) under Germany's Excellence Strategy-EXC-2111-390814868 (MCQST). F.Z. and T.X. acknowledge support from the Army Research Office under Grant No. W911NF-18-1-0416 and by the National Science Foundation under Grant Nos. DMR-1945351 through the CAREER program and DMR-1921581 through the DMREF program. **Author contributions:** F.R.G. fabricated the devices and conducted the measurements and data analysis. F.Z. and T.X. contributed the theoretical part. All authors discussed and interpreted the data. R.T.W. supervised the experiments and the analysis. The manuscript was prepared by F.R.G., F.Z. and R.T.W with input from all authors.

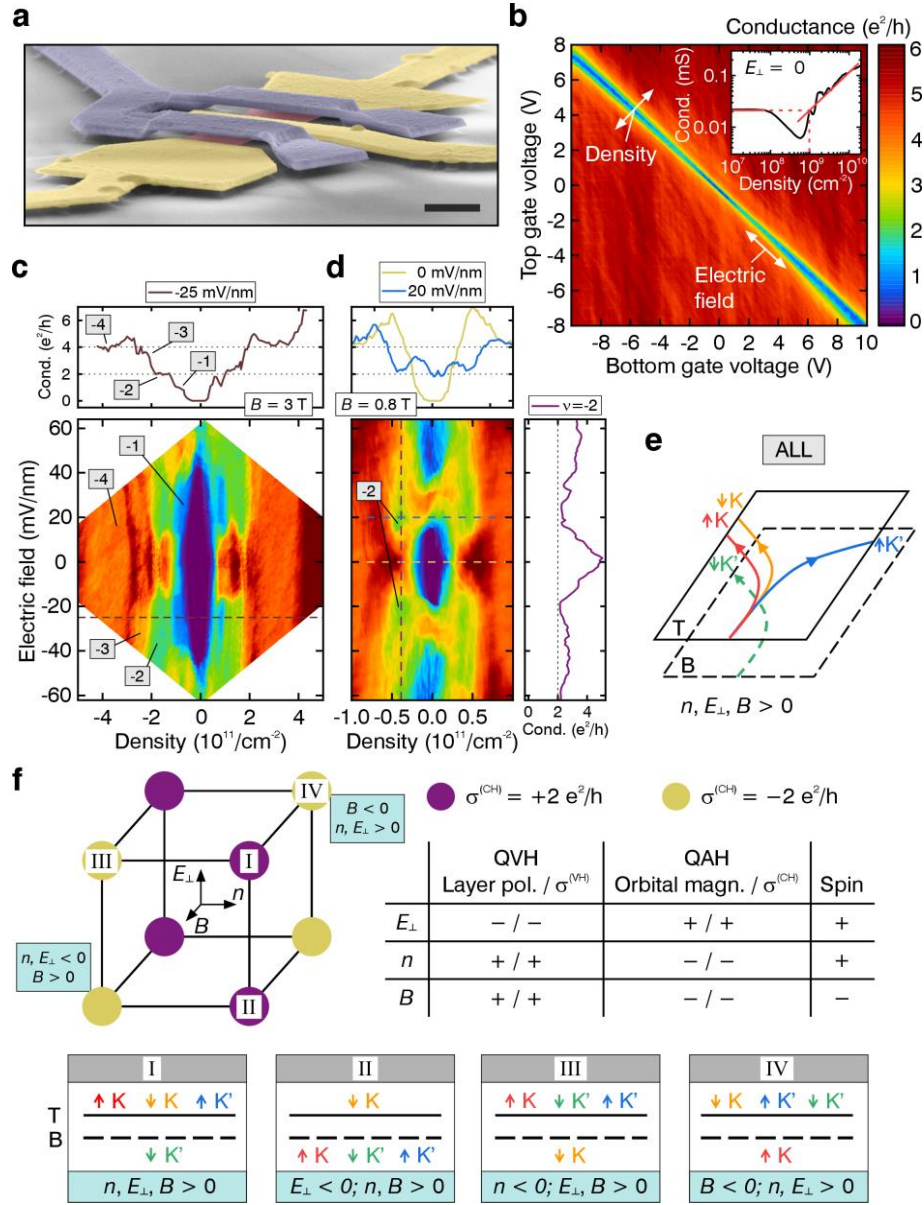

**Fig. 1. Exchange-driven quantum Hall states in dually-gated, freestanding bilayer graphene. a,** False-color scanning electron microscope image of a suspended bilayer graphene device. Contacts, top gate and bilayer graphene are shown in yellow, blue and red, respectively. Scale bar, 1 μm. **b,** Conductance map as a function of top and bottom gate voltage at $B = 0$ and $T < 10$ mK. (inset) Conductance as a function of charge carrier density at $E_\perp = 0$. The red lines are linear fits and the dashed red lines guides to the eye, indicating the residual charge carrier inhomogeneity in the device. **c,d,** Maps of the conductance as a function of $E_\perp$ and $n$ at $B = 3$ T (c) and $B = 0.8$ T (d). **e,** Schematic representation of the "ALL" quantum anomalous Hall phase in bilayer graphene showing the classical counterpart of the corresponding spontaneous quantum Hall effect for $n, E_\perp, B > 0$. T and B refer to the top and bottom

graphene layers, respectively. **f,** Top: Schematic of the eight different "ALL" phases and their corresponding Hall conductance $\sigma^{(CH)}$ and how they can be accessed by tuning $n$, $E_\perp$ and/or $B$. The table shows the properties of the QVH and QAH species of the "ALL" phase: the layer polarization and orbital magnetization as well as the valley Hall $\sigma^{(VH)}$ and charge Hall $\sigma^{(CH)}$ conductivities. $+/-$ indicates whether the observables are even/odd under flipping $n$, $E_\perp$ or $B$. Bottom: schematics of the layer polarizations of the four spin-valley species for four exemplary "ALL" phases.

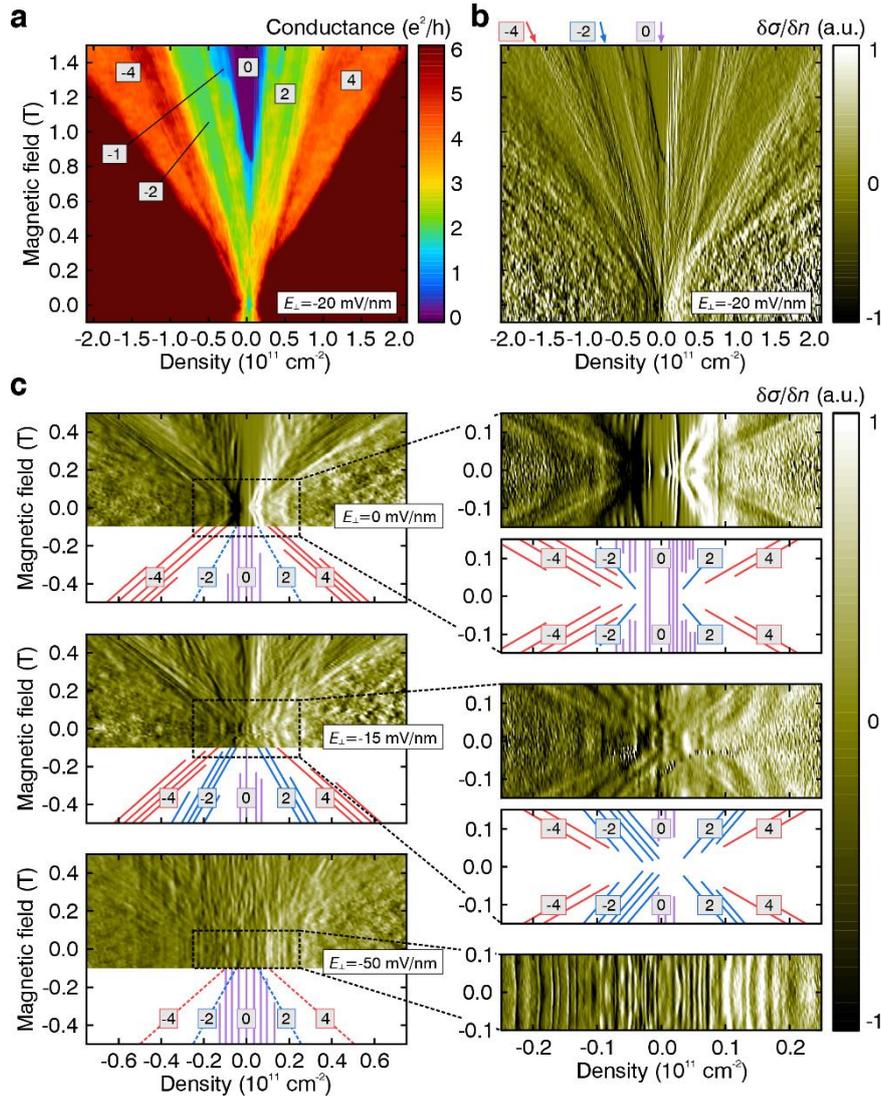

**Fig. 2. Extraordinary stability of the tunable ν = ±2 quantum Hall states towards zero magnetic field. a,b,** Fan diagram of the differential conductance (a) and its derivative $\delta\sigma/\delta n$ at $E_\perp$ = -20 mV/nm (b). The slopes of the ν = 0, -2 and -4 states are indicated with white, blue and yellow lines, respectively. **c,** Left panels: $\delta\sigma/\delta n$ plotted as a function of magnetic field and density for various $E_\perp$. Right panels: High-resolution measurement around zero magnetic field. The schematics indicate transconductance fluctuations corresponding to the ν = 0, -2 and -4 states are shown with purple, blue and red lines, respectively.

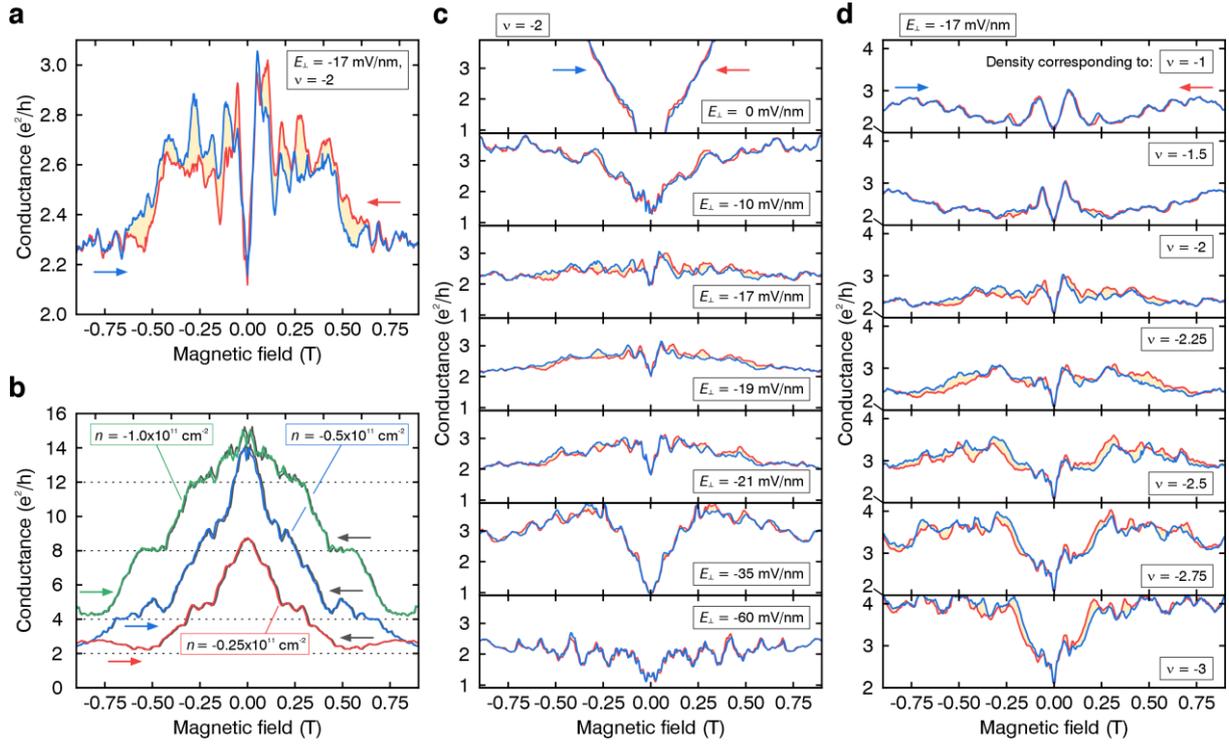

**Fig. 3. Magnetic hysteresis observable in the quantum anomalous Hall ν=-2 state in bilayer graphene. a,** Two-terminal conductance hysteresis measured for ν = -2 and $E_\perp$ = -17 mV/nm. The hysteresis loop area is shaded for clarity. **b,** Magnetic field dependence of the conductance measured for variable filling factor but fixed charge carrier density at $E_\perp$ = -17 mV/nm. The forward sweep for n = -0.25, -0.5, -1.0 × $10^{11}$ $cm^{-2}$ is shown in red, blue and green, respectively. The reverse sweeps are shown in black. **c,d,** Hysteresis of the conductance as a function of electric field at fixed ν = -2 (c) and for various filling factors at fixed $E_\perp$ = -17 mV/nm (d).

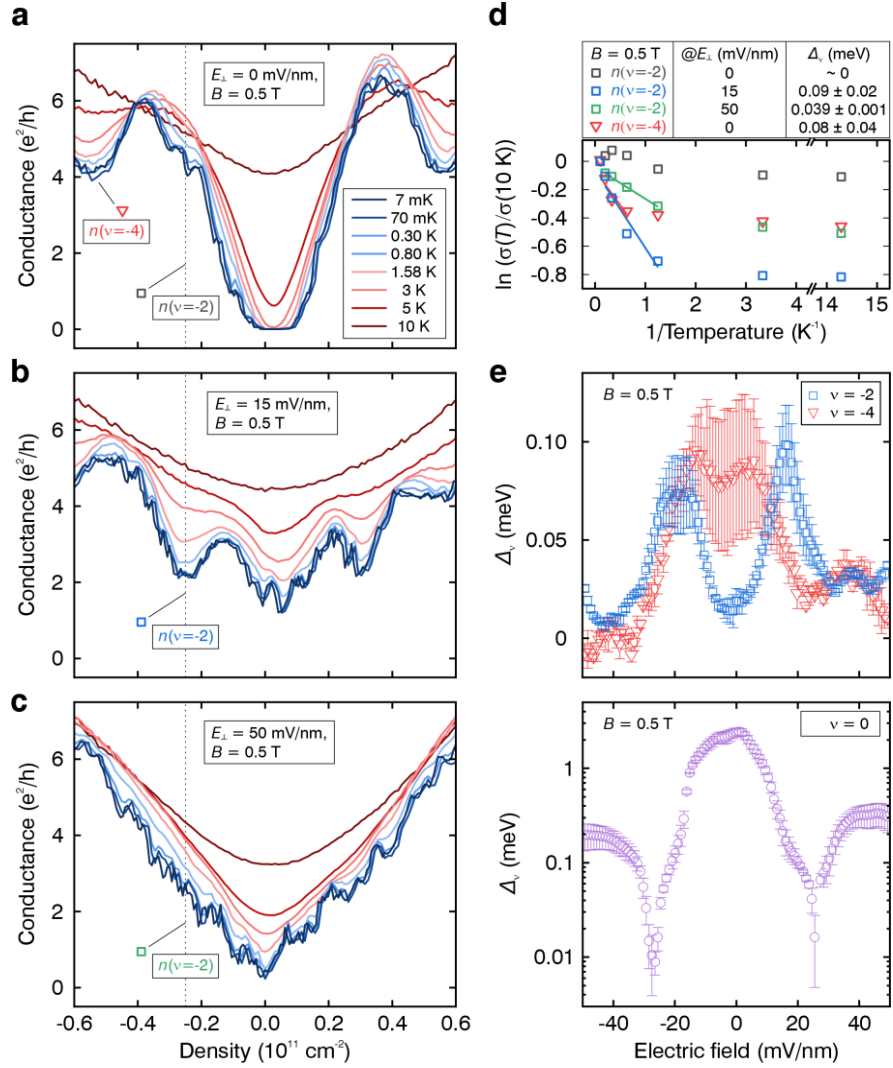

**Fig. 4. Temperature dependence of the ν = ±2 and ±4 states show distinct electric field dependence. a – c,** Conductance as a function of *n* measured for various temperatures for *B* = 0.5 T and various fixed electric fields. The density $n(\nu=-2) = -0.25 \times 10^{11}$ cm$^{-2}$ corresponding to ν = -2 at *B* = 0.5 T is indicated by the vertical lines in each plot. **d,** Arrhenius plots of the conductance measured at $n(\nu=-2)$ for $E_\perp$ = 0 mV/nm (black squares), 15 mV/nm (blue squares) and 50 mV/nm (green squares) are shown. Additionally, the data at $n(\nu=-4)$ and $E_\perp$ = 0 mV/nm is shown with red triangles. The colored lines are linear fits to the corresponding data sets. **e,** Electric field evolution of the activation gaps in the ν = 0, -2 and -4 states.

# Supplementary Materials for

## Tunable quantum anomalous Hall octet driven by orbital magnetism in bilayer graphene

**Authors:** Fabian R. Geisenhof[1], Felix Winterer[1], Anna M. Seiler[1], Jakob Lenz[1], Tianyi Xu[2], Fan Zhang[2]*, R. Thomas Weitz[1,3,4,5]*

**Affiliations:**

[1]Physics of Nanosystems, Department of Physics, Ludwig-Maximilians-Universität München, Geschwister-Scholl-Platz 1, Munich 80539, Germany

[2]Department of Physics, University of Texas at Dallas, Richardson, TX, 75080, USA

[3]Center for Nanoscience (CeNS), Schellingstrasse 4, Munich 80799, Germany

[4]Munich Center for Quantum Science and Technology (MCQST), Schellingstrasse 4, Munich 80799, Germany

[5]1st Physical Institute, Faculty of Physics, University of Göttingen, Friedrich-Hund-Platz 1, Göttingen 37077, Germany

*Corresponding authors.

Email: Zhang@utdallas.edu; thomas.weitz@uni-goettingen.de

**Current annealing procedure and calculation of the contact resistance**

Before any measurements can be performed, a current annealing procedure is used in order to clean the samples. Multiple cycles of current annealing at 1.6 K are performed, during which the DC resistance $R_{DC}$ of the sample is tracked (Fig. S1a). Generally, for an increasing applied DC voltage $V_{DC}$, the resistance of the sample decreases. However, when a saturation of the drain current is reached, $R_{DC}$ consequently increases again. The maximum current flowing was approximately 0.35 mA/$\mu$m per layer.

Since in two-terminal transport measurements there always pertains a contact resistance, we calculated and subtracted it in our data. This was done by recording a resistance versus density sweep at $B = 2$ T and $E_\perp = 20$ mV/nm. Appearing resistance plateaus were assigned to a filling factor. Plotting the resistance of the quantum Hall plateaus as a function of the inverse filling factor (Fig. S1b) gives a linear behavior. Using a linear fit demonstrates that the slope per filling factor (25604 ± 712) $\Omega$ fits well to the von Klitzing constant, while giving a contact resistance of $R_C = (3545 \pm 161)$ $\Omega$. For all measurements shown in the manuscript (except Fig. 1b), we subtracted $R_C$.

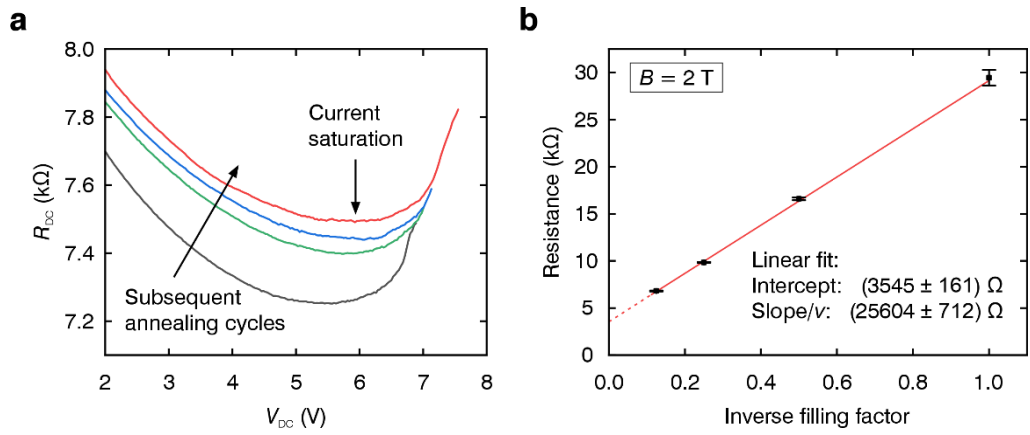

**Fig. S1. Current annealing and the extraction of the contact resistance. a,** $R_{DC}$ as a function of $V_{DC}$ during multiple annealing cycles. **b,** The resistance of quantum Hall plateaus shown as a function of inverse filling factor at $B = 2$ T and $E_\perp = 20$ mV/nm.

**Theoretical fundamentals regarding the ALL quantum anomalous Hall phase**

I. Competing ground phases in bilayer graphene at $n = E_\perp = B = 0$

In bilayer graphene at $n = E_\perp = B = 0$, when spin is ignored only two different types of competing ground phases can be distinguished[10,11]: one in which the K and K' valleys are layer polarized in the opposite sense producing a quantum anomalous Hall (QAH) phase with broken time-reversal symmetry ($\Theta$), orbital magnetization, and quantized charge Hall conductivities ($\pm 2\ e^2/h$ without counting spin degeneracy), and one in which the two valleys have the same sense of layer polarization producing a quantum valley Hall (QVH) phase with broken inversion symmetry ($P$), net layer polarization, and nontrivial valley Hall conductivity. When spin is included, there are three additional phases, namely the layer antiferromagnetic (LAF), the "ALL" as well as the quantum spin Hall (QSH) phase[10,11]. The five distinct phases in the spinful case can be obtained by each spin species choosing to be one of the two QVH phases or one of the two QAH phases, as depicted in Fig. S2. These phases are distinguished[10,11] by their charge, spin, valley, spin-valley Hall conductivities, by their orbital magnetizations as well as by their broken symmetries, as summarized in Table S1.

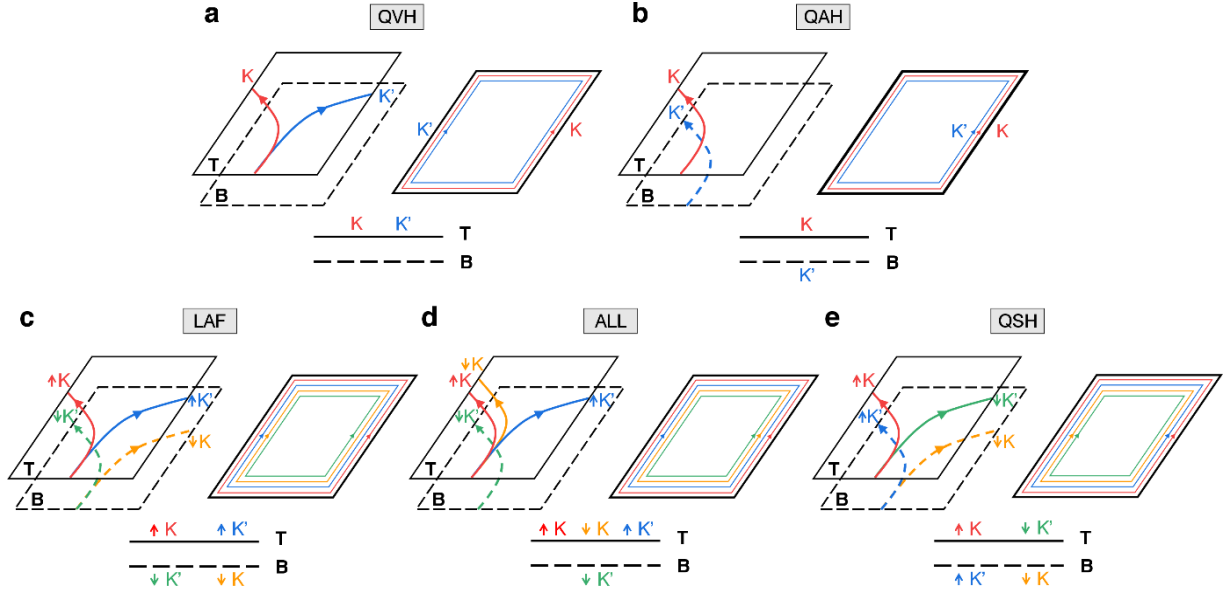

**Fig. S2. Representatives of the five competing broken symmetry ground phases in bilayer graphene at $n = E_\perp = B = 0$. a – e,** Bottom panel: layer polarization of the four spin-valley species. Top-left and top-right panels: bulk (classical) and edge (quantum) pictures of the corresponding spontaneous quantum Hall effect. Note that the edge roughness can produce couplings between counter-propagating edge states (of the same spin but different valleys) and thus gap them. Spin degeneracy is implicit in (a) and (b). See the text for details. T and B refer to the top and bottom graphene layer, respectively.

| Phase | K↑ | K↓ | K′↑ | K′↓ | Broken symm. | Mass ($\lambda\tau_z$) | $\sigma^{(CH)}$ | $\sigma^{(SH)}$ | $\sigma^{(VH)}$ | $\sigma^{(SVH)}$ |
|---|---|---|---|---|---|---|---|---|---|---|
| QVH | T | T | T | T | $P$ | $m\sigma_z$ | 0 | 0 | 2N | 0 |
| QAH | T | T | B | B | $\Theta$ | $m\tau_z\sigma_z$ | 2N | 0 | 0 | 0 |
| LAF | T | B | T | B | $\Theta, P, SU(2)$ | $ms_z\sigma_z$ | 0 | 0 | 0 | 2N |
| QSH | T | B | B | T | $SU(2)$ | $m\tau_z s_z\sigma_z$ | 0 | 2N | 0 | 0 |
| "ALL" | T | T | T | B | $\Theta, P, SU(2)$ | $m\left(\frac{1+\tau_z}{2} + \frac{1-\tau_z}{2}s_z\right)\sigma_z$ | N | N | N | N |

**Table. S1. Classification of the five competing broken symmetry ground phases in bilayer graphene at $n = E_\perp = B = 0$.** These phases are distinguished by their spin-

valley layer polarizations, by the symmetries they break, by order parameters, and by charge Hall (CH), spin Hall (SH), valley Hall (VH), and spin-valley Hall (SVH) conductivities. The data is valid in general for ABC-stacked *N*-layer graphene[10,11,18] and with *N* = 2 for AB bilayer graphene.

II. Quasiparticle orbital magnetism in bilayer graphene

In ABC-stacked *N*-layer graphene, the presence of a spontaneous gap at the Brillouin zone corners K and K' produces nontrivial momentum-space Berry curvature, and the Berry curvature gives rise to nontrivial orbital magnetic moments of quasiparticles. The orbital magnetic moment of the quasiparticle state in band α of spin $s_z$, valley $\tau_z$, and momentum $p$ reads[10,11]

$$m_{\hat{z}}^{(\alpha)}(\boldsymbol{p}, \tau_z, s_z) = \left[ -\tau_z \frac{\lambda}{h_t^2} \left( \frac{\partial h_\parallel}{\partial p} \right)^2 m_e \right] \mu_B ,$$

where $h_\parallel = (v_0 p)^N / \gamma_1^{N-1}$, $h_t = (\lambda^2 + h_\parallel^2)^{\frac{1}{2}}$, $\gamma_1 \sim 0.4$ eV is the nearest-neighbor interlayer coupling, $m_e$ is the electron mass, $\lambda \tau_z$ is the spontaneous gap term in Table S1, $\alpha = \pm$ denote the two low-energy bands, and $\mu_B$ is the Bohr magneton. Note that in the presence of a particle-hole symmetry, the moments of the particle- and hole-states are the same; in other words, the orbital magnetic moment does not depend on the band index $\alpha$.

For AB bilayer graphene, the orbital magnetic moment reads

$$m_{\hat{z}}^{(\alpha)}(\boldsymbol{p}, \tau_z, s_z) = \left[ -\tau_z \frac{4 \lambda m_e v_0^4 p^2}{\lambda^2 \gamma_1^2 + v_0^4 p^4} \right] \mu_B .$$

With $\left| W^{(\alpha)}(\boldsymbol{R}) \right\rangle = N^{-1/2} \sum_{\boldsymbol{k}} \left| \psi^{(\alpha)}(\boldsymbol{k}) \right\rangle e^{i \boldsymbol{k} \cdot \boldsymbol{R}}$ and an energy cutoff $v_0 p \sim \gamma_1 \gg |\lambda|$, it follows that the total orbital magnetization per unit area can be defined as

$$\left\langle W^{(\alpha)}(\boldsymbol{R})\left|m_{\hat{z}}^{(\alpha)}\right|W^{(\alpha)}(\boldsymbol{R})\right\rangle/A_{\text{unit-cell}} = \int m_{\hat{z}}^{(\alpha)}(\boldsymbol{p},\tau_z,s_z)\frac{d\boldsymbol{k}}{(2\pi)^2} = -\tau_z\frac{\lambda\, m_e}{\pi\, \hbar^2}\ln\left(\frac{\gamma_1}{|\lambda|}\right)\mu_B\,.$$

For a spontaneous gap of 10, 1, and 0.1 meV, the orbital magnetization per unit cell for each spin-valley species is 8.0, 1.3, and 0.18 $m\mu_B$, respectively.

III. Eight possible ALL quantum anomalous Hall phases in bilayer graphene

    The ALL phase in Table S1 and Fig. S2 can be viewed as a phase in which one spin-valley species polarizes into one layer whereas the other three species polarize into the opposite layer, or alternatively as a phase in which one spin species is in one of the two possible QVH phases (that have opposite layer polarization, e.g., Figs. S3a and b) whereas the other spin species is in one of the two possible QAH phases (that have opposite Chern numbers, e.g., Figs. S3a and f). Based on either viewpoint, one can find eight different ALL phases in total, as depicted in Fig. S3.

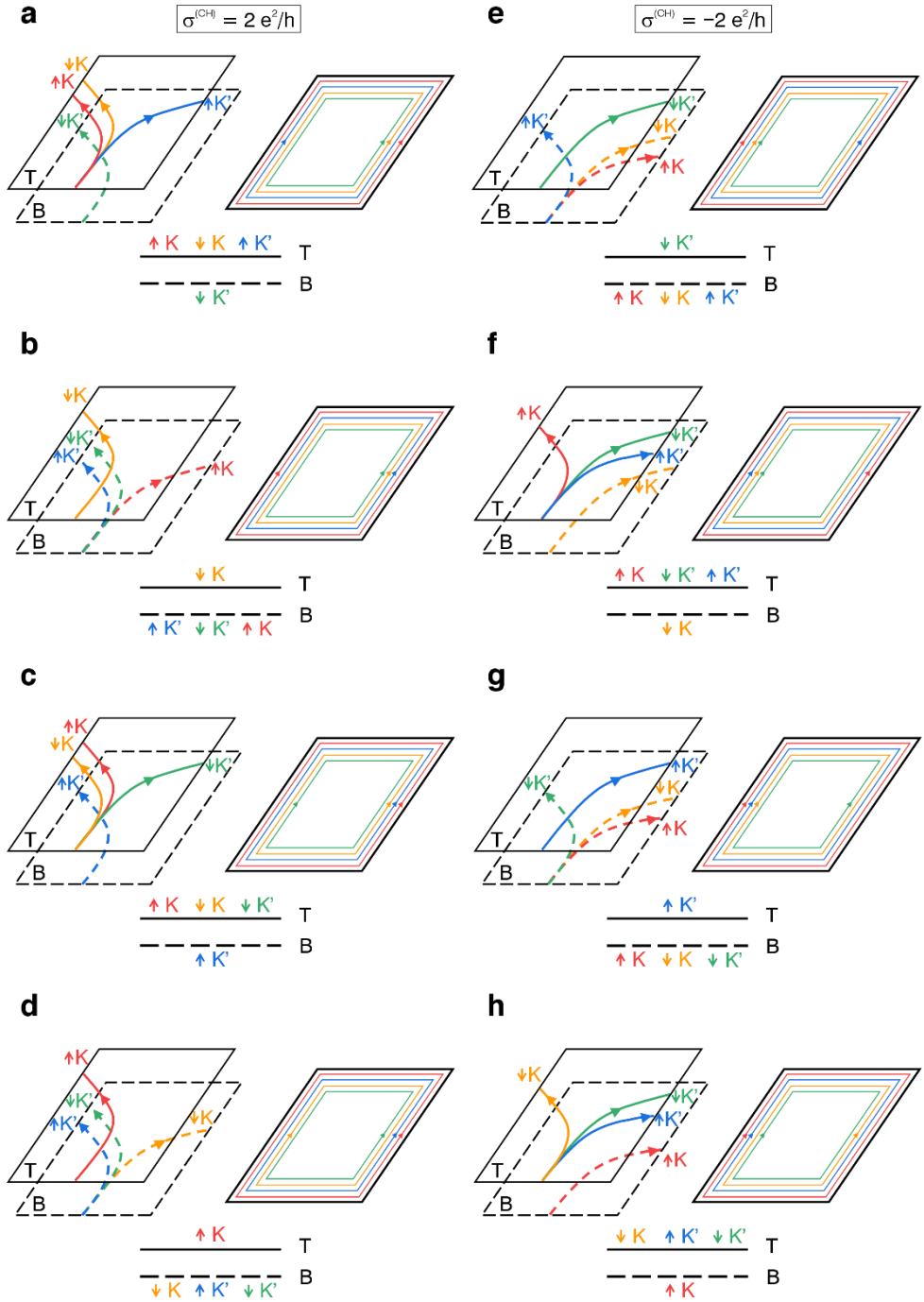

**Fig. S3**. **Possible ALL quantum anomalous Hall phases in bilayer graphene. a – h** Eight different ALL phases that can be classified by the polarizations of their two spin species, by which spin species being in which QAH or QVH phases, and by their charge Hall conductivities.

**Evolution of the $\nu = \pm 2$ state in electric and magnetic field**

Here we show additional data on how the $\nu = \pm 2$ state behaves in an electric and magnetic field. We have recorded multiple electric field versus density conductance maps at various magnetic fields (Fig. S4a – h). Figure S4a – d shows the conductance map for lower magnetic fields $B$ = 0.1 T, 0.2 T and 0.5 T as well as a map with a reversed field of $B$ = -0.5 T, respectively. Of the four domains observed at $B$ = 0.8 T, only three show a quantized conductance of 2 $e^2$/h at lower fields. The domain at negative electric field and positive density shows a higher conductance, possibly due to residual disorder providing additional channels for charge transport. Still, this domain behaves like the other three, as we also see in the fan diagrams in Fig. 2 in the main text.

Changing the direction of the magnetic field (Fig. S4c and d) shows the other four ALL phases (see also Fig. S3).

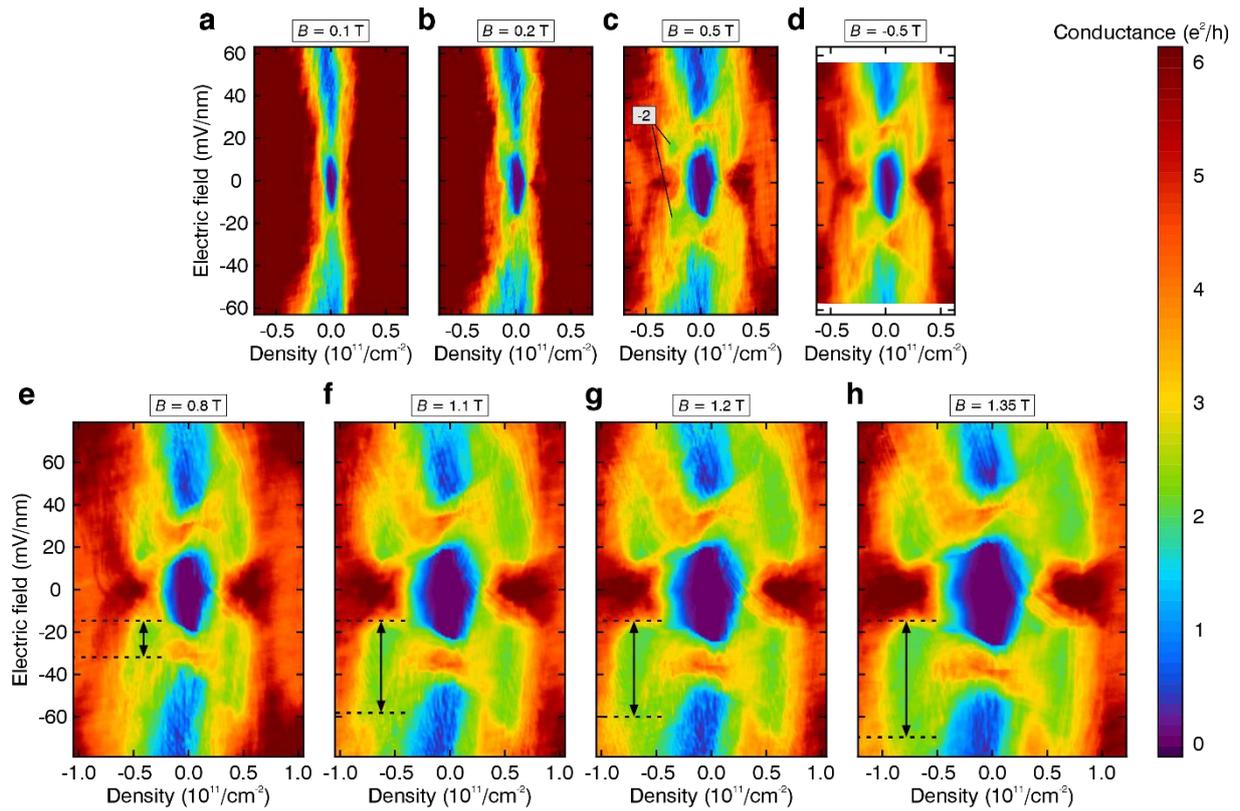

**Fig. S4. Additional measurements showing the electric and magnetic field dependence of the ν = ±2 state. a – h,** Maps of the conductance as a function of electric field and density for various magnetic fields. The dashed lines in (e – h) are guides to the eye, and the arrows indicate the range of negative electric field at which the ν = -2 state emerges.

**Additional fan diagrams showing a complete electric field series**

Figure S5 shows additional fan diagrams, demonstrating the behavior of quantum Hall states towards zero magnetic field for various electric fields. The strength of each Landau level is indicated by the amount of colored lines with the corresponding slope in the top of each picture.

Since the $\nu = \pm 4$ state is a non-layer polarized phase, it is less and less pronounced for increasing electric field. On the contrary, as discussed already in the main manuscript, the $\nu = \pm 2$ state is strongest for a finite range of electric fields. However, it does already emerges at $E_\perp = 0$ but disappears for increasing magnetic fields. For $E_\perp = -10$ mV/nm it does finally emerge for the complete magnetic field range shown here. The highest amount of fluctuations corresponding to it appears at $E_\perp = -15$ mV/nm to -20 mV/nm, while for higher negative fields they disappear again. Lastly, the $\nu = 0$ state is strong for low electric fields (canted antiferromagnetic phase), and for very high electric fields, where it is a fully layer polarized phase.

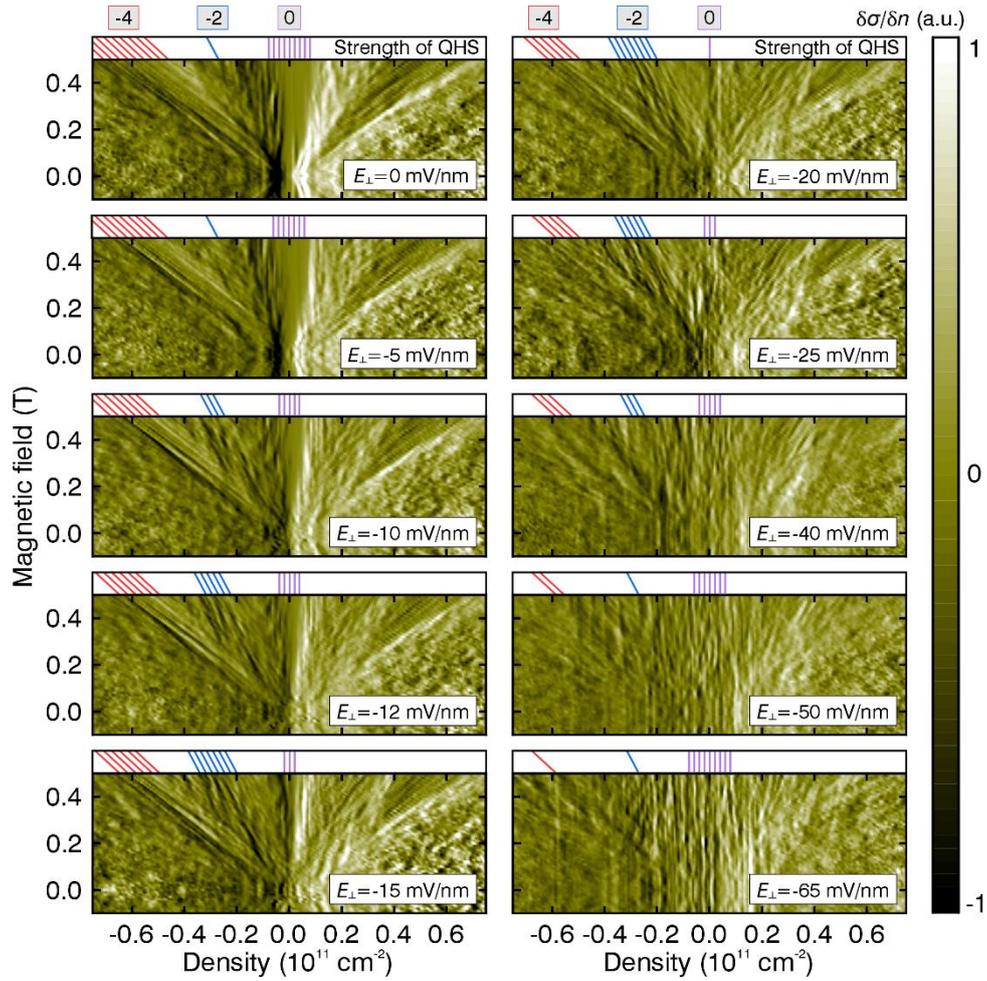

**Fig. S5. Tracing quantum Hall states towards zero magnetic field.** Derivative of the differential conductance δσ/δn plotted as a function of magnetic field and density for various $E_\perp$. The amount of conductance fluctuations corresponding to the $\nu$ = 0, -2 and -4 state are indicated by the number of white, blue and yellow lines in the top of each image.

**Temperature dependent transport data**

Figure S6 shows maps of the conductance as a function of electric field and density for various temperatures. The temperature dependent data shown in Figure 4 in the main manuscript are taken from these measurements, with the position of the linecuts indicated by dashed lines in the top left image. In general, we see that for $T \leq 0.3$ K the maps are basically the same, while for higher temperatures the $\nu = \pm 2$ and $\pm 4$ as well as the $\nu = 0$ state get less and less well resolved, since fluctuations due to increasing temperatures broaden all phase transitions. We would like to point out, that gap energies measured by activation can only give a lower bound for the real gap due to the presence of local disorder. As consequence, in some measurements (e.g. Chen *et al.* Ref. [8]) activation gaps are - like in our case - smaller than the temperature range they are measured in. To give an estimate of thermodynamic gaps, direct measurements of the inverse compressibility would be required (Feldman *et al.* Ref. [30]). Our gap energies should be thus understood as lower bound and can give an estimate to compare the strength of the different phases against one another within the same sample and to get a feeling for the dependence of the gap strength as function of applied perpendicular electric field.

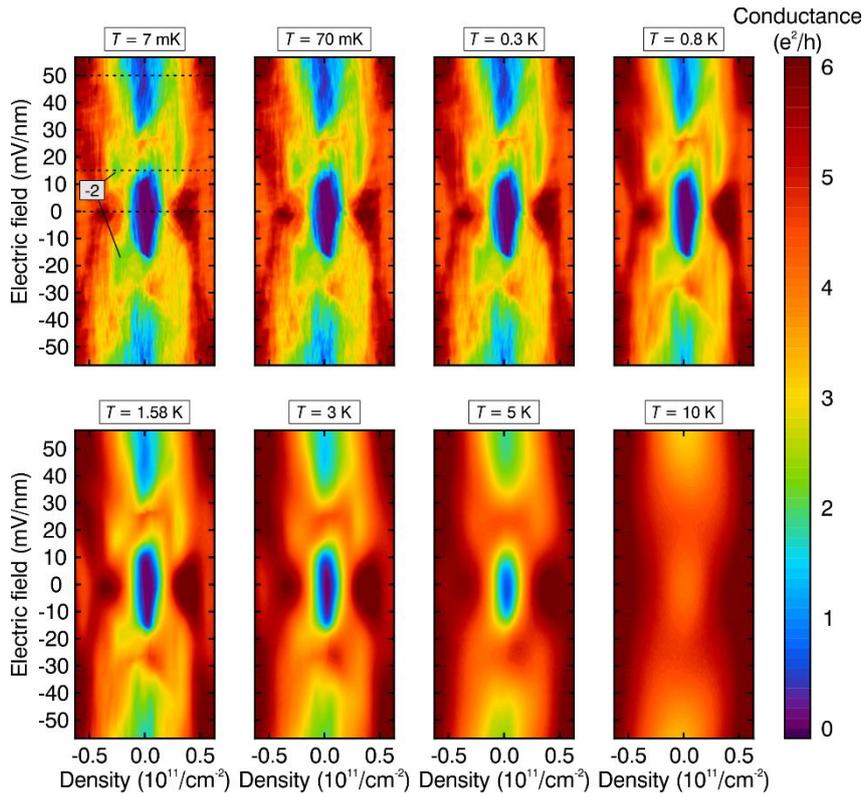

**Fig. S6. Temperature dependence of the quantum Hall states at $B$ = 0.5 T.** Map of the conductance as a function of density and electric field for various temperatures. The dashed lines in the top left images indicate the position of the linecuts shown in Fig. 4 in the main manuscript.

**Evidence of the quantum anomalous Hall effect in a second device.**

Figure S7 shows the quantum transport data measured in a second device. Figure S7a,b show the conductance map for low magnetic fields of $B$ = 0.2 T and 0.5 T, respectively. While the sample is less clean than the one shown in the main manuscript, we still see four domains of with conductance $\pm 2\ e^2/h$ (four green-colored regions in Fig. S7a,b even at these low magnetic fields. Furthermore, the $\nu = \pm 2$ states have the same behavior when applying an electric field and magnetic field. Fig. S7c shows the conductance as a function of electric and magnetic field for a fixed filling factor $\nu$ = -2. The $\nu$ = -2 state does only emerge for intermediate applied electric fields and the range at which it appears increases with increasing magnetic field. Lastly, also in this device we see magnetic hysteresis when sweeping B around zero while fixing $\nu$ = -2 and $E_\perp$ = -19 mV/nm. However, the hysteresis is less prominent and the conductance breaks down for low magnetic fields, presumably due to the lower quality of the device.

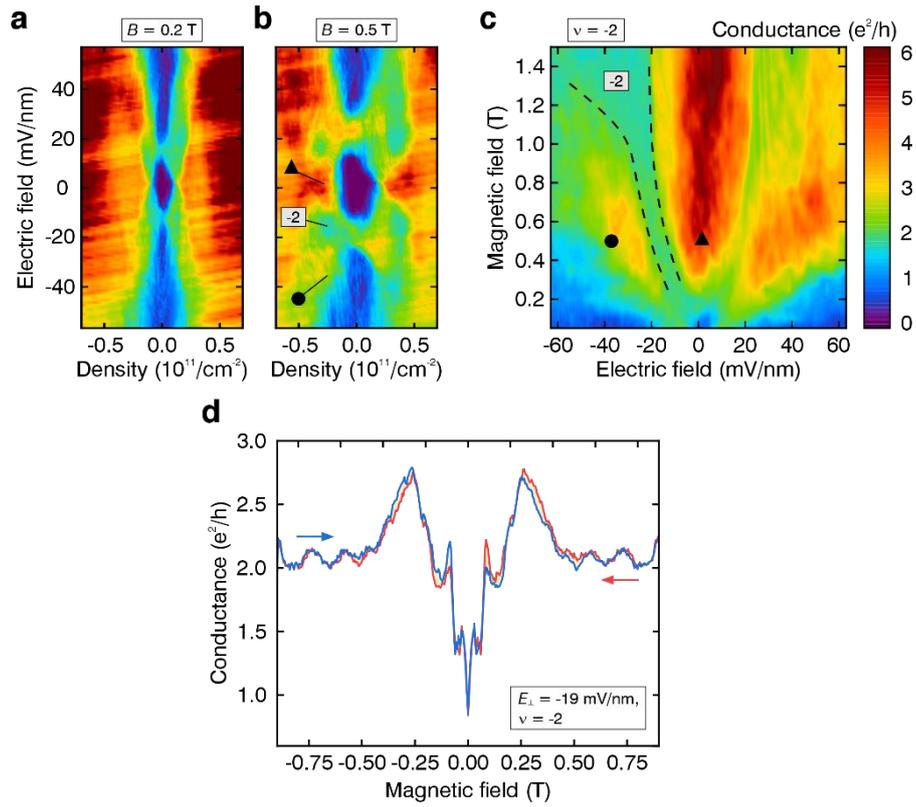

**Fig. S7. Quantum transport data in a second device. a,b,** Maps of the conductance as a function of $E_\perp$ and $n$ for $B = 0.2$ T and 0.5 T, respectively. **c,** Conductance as a function of electric and magnetic field for fixed filling factor of $\nu = -2$. The dashed lines indicate the region where the $\nu = -2$ state at negative electric fields emerges with a conductance of 2 $e^2/h$. **d,** Two-terminal conductance hysteresis measured for $\nu = -2$ and $E_\perp = -19$ mV/nm. The hysteresis loop area is shaded for clarity.

# References

1. Zhang, F. Spontaneous chiral symmetry breaking in bilayer graphene. *Synth. Met.* **210,** 9–18 (2015).

2. Nandkishore, R. & Levitov, L. Quantum anomalous Hall state in bilayer graphene. *Phys. Rev. B* **82** (2010).

3. Shi, Y. *et al.* Electronic phase separation in multilayer rhombohedral graphite. *Nature* **584,** 210–214 (2020).

4. Chen, G. *et al.* Tunable correlated Chern insulator and ferromagnetism in a moiré superlattice. *Nature* **579,** 56–61 (2020).

5. Martin, J., Feldman, B. E., Weitz, R. T., Allen, M. T. & Yacoby, A. Local Compressibility Measurements of Correlated States in Suspended Bilayer Graphene. *Phys. Rev. Lett.* **105,** 256806 (2010).